\documentclass[aps,pra,twocolumn,showpacs,10pt]{revtex4}
\usepackage{color, graphicx}
\usepackage{amsmath, amssymb}
\usepackage{ulem,hyperref}

\begin{document}

\title{Unified quantification of nonclassicality and entanglement}
\author{W. Vogel}\affiliation{Arbeitsgruppe Theoretische Quantenoptik, Institut f\"ur Physik, Universit\"at Rostock, D-18051 Rostock, Germany}
\author{J. Sperling}\affiliation{Arbeitsgruppe Theoretische Quantenoptik, Institut f\"ur Physik, Universit\"at Rostock, D-18051 Rostock, Germany}
\date{\today}

\begin{abstract}
	The nonclassicality of single-mode quantum states is studied in relation to the entanglement created by a beam splitter.
	It is shown that properly defined quantifications -- based on the quantum superposition principle -- of the amounts of nonclassicality and entanglement are strictly related to each other.
	This can be generalized to the amount of genuine multipartite entanglement, created from a nonclassical state by an $N$ splitter.
	As a consequence, a single-mode state of a given amount of nonclassicality is fully equivalent, as a resource, to exactly the same amount of entanglement.
	This relation is also considered in the context of multipartite entanglement and multimode nonclassicality.
\end{abstract}
\maketitle

\section{Introduction}
Nonclassical properties of light became a subject of increasing interest, stimulated by the availability of the coherent light sources since the 1960s.
This also led to new developments of the quantum coherence theory of light~\cite{glauber1,glauber2,glauber3,klaud-sud}.
The properties of coherent states appear to be most close to the classical behavior of a light field~\cite{tit-glau,mandel,ma-wo,vo-we,agar}.
For notational simplicity, we denote them as classical states, with respect to the property nonclassicality.
Any quantum state $\hat \rho$ can be represented in terms of the Glauber-Sudarshan $P$~representation~\cite{glauber2,sud},
\begin{equation}
  \hat \rho = \int {\rm d} P(\alpha) |\alpha \rangle \langle \alpha|,
  \label{eq:noncl}
\end{equation}
which resembles a mixture of classical (i.e., coherent) states $|\alpha \rangle$.
Whenever the $P$~function, $P(\alpha)$, can be interpreted as a classical probability density, the quantum state is a true mixture of classical states and, hence, it is called classical.
If such a representation does not exist, i.e., $P(\alpha)\not\ge 0$, the quantum state is called nonclassical; see, e.g., Refs.~\cite{tit-glau,mandel,ma-wo,vo-we,agar}.
Any nonclassical state includes quantum superpositions of coherent states.

It is interesting that the notion of entanglement developed almost independently from that of nonclassicality. 
The first studies of entanglement date back to 1935, focusing on surprising consequences of the quantum description of nature~\cite{epr,schroed}. 
Nowadays, entanglement is considered to be a fundamental resource for quantum information, quantum computation, quantum metrology, and other applications -- altogether denoted as quantum technology~\cite{nielsenchuang,horodecki,guehne}.
To characterize the property entanglement, let us consider a bipartite quantum state,
\begin{equation}
  \hat \rho = \int {\rm d} P(a,b) |a,b \rangle \langle a,b|,
  \label{eq:ent}
\end{equation}
with the notation $|a,b\rangle =|a\rangle \otimes |b\rangle$ for product states in the Hilbert spaces of the parties A and B, $|a\rangle \in \mathcal{H}_\text{A},~|b\rangle \in \mathcal{H}_\text{B}$.
If $P(a,b)$ is a classical probability, the state is called separable~\cite{werner}.
If such a representation cannot be found, i.e., $P(a,b)\not\ge 0$, the state is entangled~\cite{sanpera,spe-vo2}.
In this case the considered state requires the global quantum superposition of product states of both parties.
Similar to the property nonclassicality, a product state $|a,b\rangle$ may be considered to be classical, now with respect to the property entanglement. 

The structures of nonclassical and entangled quantum states, Eqs.~\eqref{eq:noncl} and~\eqref{eq:ent}, are very similar in a formal sense.
The origin of the quantum effects we are interested in is the quantum superposition principle.
In the words of Ref.~\cite{Raymond}:
''The superposition principle is at the heart of the most intriguing features of the microscopic world.''
Correspondingly, these authors stated: ''When the superposition principle is 	applied to composite systems, it leads to the essential concept of entanglement.''
On this basis the question arises: To which extent can the property nonclassicality of a single-mode system be related to bipartite entanglement?
Most importantly, does such a formal equivalence imply that nonclassical single-mode states can be considered as the resource for practical applications, which usually require bipartite entanglement?

In the present paper we prove the close relation bet\-ween the quantification of nonclassicality and entanglement.
A unitary transformation, as it is realized by an optical beam splitter, was known to convert nonclassicality of a single-mode radiation field into bipartite entanglement~\cite{ahar,kim,xiangbin,WEP03,JLC13}.
Beyond this fact, we show that the amount of nonclassicality of a single-mode radiation field is strictly transformed into the same amount of bipartite entanglement.
More generally, the available amount of nonclassicality can even be converted into the same amount of genuine multipartite entanglement.
Altogether, this implies that any amount of entanglement desired as a resource for applications in quantum technology is equivalent to the same amount of nonclassicality of a single-mode state.

The paper is structured as follows.
In Sec.~\ref{sec:uq} we introduce the unified quantification of nonclassicality and entanglement on the basis of the quantum superposition principle.
Some elementary examples of the mapping of nonclassicality onto entanglement are considered in Sec.~\ref{sec:ee}.
Section~\ref{sec:gqr} is devoted to the derivation of the general quantitative relation between nonclassicality and entanglement even in the multi-mode case.
A summary and some conclusions are given in Sec.~\ref{sec:sc}.

\section{Unified quantification}\label{sec:uq}
To quantify the properties nonclassicality and entanglement, a number of different approaches exists.
We restrict attention here to those approaches which are relevant in our context.
Of some interest is a proposal to quantify nonclassicality through the so-called entanglement potential~\cite{asboth}.
The problem of this idea is that a manifold of entanglement measures was considered in the literature; cf., e.g., Refs.~\cite{horodecki,guehne}.
Of course, different entanglement measures may lead to different quantifications of the nonclassicality.

Our approach is fundamentally different in this respect.
We first fix the quantifications of nonclassicality and entanglement, in a manner that relies on the basic algebraic structures of these kind of quantum states, in close relation to the fundamental quantum superposition principle.
In the next step, we study the consequences for the quantitative relation of nonclassicality and entanglement.
For quantifying entanglement, the Schmidt number of, in general, mixed quantum states was known to obey the requirements of a measure; cf., e.g., Ref.~\cite{horodecki}.
Let a given pure, bipartite state $|\Psi_{\rm Ent}\rangle$ have a Schmidt decomposition~\cite{nielsenchuang},
\begin{equation}
 |\Psi_{\rm Ent}\rangle = \sum_{i=1}^r \lambda_i |a_i,b_i\rangle,
 \label{eq:Schmidt-dec}
\end{equation}
with the positive Schmidt coefficients $\lambda_i$ and the pairwise orthonormal states $|a_i\rangle \in \mathcal{H}_\text{A},~|b_i\rangle \in \mathcal{H}_\text{B}$.
This state is more entangled as its Schmidt rank $r$ is larger, which counts to which extent the state includes quantum superpositions of product states.
Note that the number $r$ of superpositions is not only relevant as a fundamental concept of quantum physics but also for practical applications.
For example, the accessible alphabet for quantum communication and computation is increased with increasing $r$ values.
This is relevant for the teleportation of high-dimensional quantum states~\cite{JJS09} or security against eavesdropping~\cite{Z11}.
Additionally, the complexity of possible processes to be handled by a local unitary operation is also growing.

For quantifying nonclassicality, one may proceed in a formally similar way.
Let a pure single-mode state $|\Psi_{\rm Ncl}\rangle$ be expressed in terms of superpositions of $r$ coherent states $|\alpha_i\rangle$,
\begin{equation}
 |\Psi_{\rm Ncl}\rangle = \sum_{i=1}^r \kappa_i |\alpha_i\rangle,
 \label{eq:CS-dec}
\end{equation}
with $\alpha_i \not= \alpha_j$ for $i\not=j$, and complex values of $\kappa_i\neq0$.
This minimal number $r$ of superpositions defines the nonclassicality measure of the state~\cite{gehrke}; it counts the desired number of superpositions of classical states $|\alpha_i\rangle$.

The quantification of entanglement and nonclassicality by a number $r$ has common features from the fundamental point of view.
First, in both cases the number $r$ quantifies the extent to which we make use of the quantum superposition principle to represent the given state.
Second, the extension of this quantification to mixed states is based on a convex roof construction~\cite{uhlmann}, leading to the Schmidt number as an entanglement quantifier.
The same can be done for mixed single-mode nonclassicality.
Third, the Schmidt number and its monotones have been proven to be the only universal entanglement measures; i.e., they do not increase under all separable operations~\cite{spe-vo1}.
This implies the impossibility of increasing this entanglement measure by any local operation.
Correspondingly, both the number of superpositions of coherent states and its monotones do not increase under all classical operations~\cite{gehrke}.

Beside these similarities, however, there exists a mathematical difference between the representations of $|\Psi_{\rm Ent}\rangle$ and $|\Psi_{\rm Ncl}\rangle$ and the respective quantifications.
Whereas the former state is expressed in terms of orthonormal states, the latter is not, since the coherent states are non-orthogonal.
Hence, we have to study the consequences of this difference in more detail.
Before doing this, we consider some elementary examples.

\section{Elementary examples}\label{sec:ee}
\begin{figure}[h]
\includegraphics*[width=4.5cm]{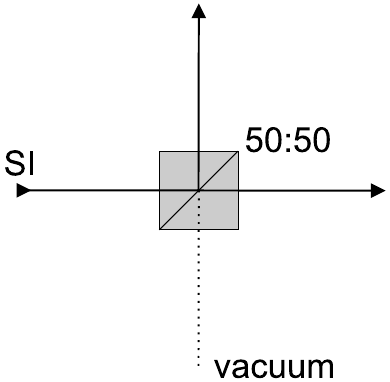}
\caption{
	A classical or nonclassical field is prepared.
	The signal field (SI) is combined on a $50:50$ beam splitter (BS) with vacuum.
	Depending on the input state, the output can be highly or weakly entangled, or even separable.
}\label{fig:setup}
\end{figure}

Let us consider the action of a symmetric $50:50$ beam splitter, producing typical entangled quantum states out of a nonclassical state and a vacuum input state; see Fig.~\ref{fig:setup}.
We start with fields in the input ports to be the coherent state $|\alpha\rangle$ and the vacuum state $|0\rangle$.
In the output ports we get a product of coherent states,
\begin{equation}
 |\alpha,0\rangle \,\stackrel{\rm BS}{\longmapsto}\, \left|\frac{\alpha}{\sqrt 2},\frac{\alpha}{\sqrt 2}\right\rangle,
 \label{eq:CS-BS}
\end{equation}
by selecting, without loss of generality, a phase difference in the two output ports. 
It is well known that any classical input yields a separable output for the scenario in Fig.~\ref{fig:setup}; cf.~\cite{ahar,kim,xiangbin,WEP03,JLC13}.

Now we consider an example of a weakly nonclassical input state, i.e., $r=2$, the so-called odd coherent state~\cite{dodonov}, in one of the input channels.
The action of the beam splitter yields
\begin{align}
 &\mathcal N_\alpha\left(|\alpha\rangle - |-\alpha\rangle\right)\otimes|0\rangle\ 
 \nonumber
 \\ \,\stackrel{\rm BS}{\longmapsto}\,
 &\mathcal N_\alpha\left(\left|\frac{\alpha}{\sqrt{2}},\frac{\alpha}{\sqrt{2}}\right\rangle - \left|-\frac{\alpha}{\sqrt{2}},-\frac{\alpha}{\sqrt{2}}\right\rangle\right),
 \label{eq:ent-ECS}
\end{align}
with $\mathcal N_{\alpha}=\left[2\left(1-\exp[-2|\alpha|^2]\right)\right]^{-1/2}$.
The resulting quantum state in the output ports is clearly entangled.
In the representation with coherent states it is a superposition of two product states, resulting from the properties of the nonclassical input state. 

It is interesting to consider this result in the limit of vanishing coherent amplitude, $|\alpha|\to 0$.
In this case, Eq.~\eqref{eq:ent-ECS} reduces -- up to a global phase -- to
\begin{equation}
 |1,0\rangle  \,\stackrel{\rm BS}{\longmapsto}\, \frac{1}{\sqrt{2}}(|1,0\rangle + |0, 1\rangle),
 \label{eq:ent-photon}
\end{equation}
so that a single photon transforms into a Bell state.
The expression on the right-hand side is given in the Schmidt decomposition, indicating a weak entanglement with a Schmidt rank of only $r=2$.
This holds also true for the expression~\eqref{eq:ent-ECS} for general values of $\alpha$, as we clearly demonstrate below.

Let us consider another example with the inputs being a vacuum state and a squeezed vacuum state~$|\text{sv}\rangle$.
The latter can be expressed as
\begin{equation}
 |\text{sv}\rangle = \frac{1}{\sqrt{\mu}} \exp\left[-\frac{\nu}{2\mu}\hat a^{\dagger\,2}\right] |0\rangle,
\end{equation}
with the real and complex parameters $\mu$ and $\nu$, respectively; cf., e.g., Ref.~\cite{vo-we}.
The parameter $|\nu|=\sqrt{\mu^2 -1}$ controls the quantum noise suppression of the squeezed state.
The latter may be expanded in terms of coherent states as
\begin{equation}
 |\text{sv}\rangle = \int \frac{{\rm d}^2 \alpha}{\pi\sqrt{\mu}} \exp\left[-\frac{\nu}{2\mu} \alpha^{\ast\,2}-\frac{|\alpha|^2}{2}\right] |\alpha\rangle.
\end{equation}
Making use of Eq.~\eqref{eq:CS-BS}, we get in the output channels of the beam splitter the entangled state
\begin{equation}
 |\text{sv},0\rangle \,\stackrel{\rm BS}{\longmapsto}\, \int \frac{{\rm d}^2 \alpha}{\pi\sqrt{\mu}} \exp\left[-\frac{\nu}{2\mu} \alpha^{\ast\,2}-\frac{|\alpha|^2}{2}\right] |\frac{\alpha}{\sqrt{2}},\frac{\alpha}{\sqrt{2}}\rangle.
\end{equation}
The input state $|\text{sv}\rangle$ is necessarily expressed by an infinite number of superpositions of coherent states~\cite{gehrke}.
The representation of the entangled output state requires an infinite number of superpositions of products of coherent states.
Hence, a strongly nonclassical input state, with $r=\infty$, yields a correspondingly strong entanglement in the output channels.

At this point it is worth commenting on the quantification of nonclassical states, as considered recently in Ref.~\cite{gehrke}.
In this paper we had introduced an algebraic approach of quantifying the state under study by the number of superpositions of coherent states required for its representation.
The conclusion was that the representation of the Fock state requires an infinite number of coherent states.
This holds true, as long as we consider representations excluding the degenerate case, in which some of the coherent states may become equal.
In the example of splitting a single photon by a beam splitter, we have exactly such a degenerate situation.
It can be seen from Eqs.~\eqref{eq:ent-ECS} and~\eqref{eq:ent-photon} that two coherent states are enough to represent the single photon state, if we include the degenerate representation -- contrary to the approach in Ref.~\cite{gehrke}.

In a similar way, cf. Eqs.~\eqref{eq:ent-ECS} and~\eqref{eq:ent-photon}, we can represent any Fock state $|n\rangle$ as the limit of a difference quotient:
\begin{align}
 \nonumber |n\rangle =& \frac{\hat a^\dagger{}^n}{\sqrt{n!}}|0\rangle=\left.\partial_\alpha^n \left(\frac{1}{\sqrt{n!}}\exp\left[\frac{|\alpha|^2}{2}\right]|\alpha\rangle\right)\right|_{\alpha=0}
 \\=& \lim_{\alpha\to 0} \frac{\sum_{j=0}^n \frac{n!(-1)^{n-j}}{j!(n-j)!} \exp\left[\frac{|j\alpha|^2}{2}\right]|j\alpha\rangle}{\sqrt{n!}|\alpha|^n},
\end{align}
which requires the superposition of $r=n+1$ coherent states.
This result is fully consistent with the splitting of an $n$-photon Fock state by the beam splitter in Fig.~\ref{fig:setup}, 
\begin{equation}
 |n,0\rangle \,\stackrel{\rm BS}{\longmapsto}\, \frac{1}{2^{n/2}}\sum_{j=0}^n \binom{k}{j}^{1/2} |j,n-j\rangle,
\end{equation}
which yields an entangled state with a Schmidt rank of $r=n+1$.

In this context it is worth remembering the Hahn-Banach separation theorem~\cite{HBBuch}, which can be applied to separate the convex set of up to $r$ superpositions from the remaining states.
Its proof requires closed sets, e.g., the closure of all convex combinations of pure states with a number of superpositions less than or equal to $r$.
On this basis, the quantification of both entanglement and nonclassicality as considered in the context of Eqs.~\eqref{eq:Schmidt-dec} and~\eqref{eq:CS-dec}, respectively, has to include this closure.
As a consequence, the nonclassicality quantification yields the result that a squeezed vacuum is strongly nonclassical, when compared with an $n$-photon Fock state.
As found above, by splitting the corresponding states with a beam splitter, we get stronger entanglement for the squeezed vacuum as for the $n$-photon state.
Let us note that the Wigner function of any photon-number state (with $n>0$) has negative contributions, whereas the Wigner function of a squeezed state is even a classical (Gaussian) one.
This implies that negativities of the Wigner function are no indication of the strength of nonclassicality of a quantum state, in the sense of the quantum superposition principle.

\section{General quantitative relation}\label{sec:gqr}
It can be shown in two steps that the setup in Fig.~\ref{fig:setup} maps the nonclassicality measure of the input to the entanglement measure of the output.
First, we show that a superposition of $r$ coherent states yields a superposition of an output state with a Schmidt rank $r$.
Second, the first step together with the fact that the beam splitter preserves the purity of any input state, i.e. its mixing properties, implies that the amount of nonclassicality of an arbitrary mixed input state transforms to exactly the same amount of entanglement in the output ports.

\subsection{Bipartite entanglement}
Let us consider the mapping of the input state $|\Psi_{\rm Ncl}\rangle$ defined in Eq.~\eqref{eq:CS-dec}.
We get
\begin{equation}
 \sum_{i=1}^r \kappa_i |\alpha_i,0\rangle
 \,\stackrel{\rm BS}{\longmapsto}\, \sum_{i=1}^r\kappa_i\left|\frac{\alpha_i}{\sqrt 2},\frac{\alpha_i}{\sqrt 2}\right\rangle.
 \label{eq:BSexpansion}
\end{equation}
Since the right-hand side is not the Schmidt decomposition, we still have to show that the Schmidt rank agrees  with the number $r$ of superimposed coherent states of the input state.
Note that the Schmidt rank $r$ is independent of the basis expansion.
If we can prove the linear independence of the coherent states, the Schmidt decomposition is readily obtained by an orthonormalization procedure.

In order to prove that Eq.~\eqref{eq:BSexpansion} is given in terms of linearly independent vectors, we have to prove that any set $\{|\alpha_1\rangle,\ldots,|\alpha_r\rangle\}$, with $\alpha_i\neq\alpha_j$ for $i\neq j$, is linearly independent.
For simplicity, we ignored the scaling $2^{-1/2}$ of the coherent amplitudes.
Let us consider the Fock basis expansion, $|\alpha\rangle=\sum_{n=0}^\infty\exp[-|\alpha|^2/2]\alpha^n|n\rangle/\sqrt{n!}$, which we truncate to photon numbers below $r$.
The resulting quadratic matrix $\boldsymbol M$ is defined by its elements
\begin{equation}
 M_{j,n}=\exp\left[-\frac{|\alpha_j|^2}{2}\right]\frac{{\alpha_j}^n}{\sqrt{n!}},
\end{equation}
for $n=0,\ldots,r-1$ and $j=1,\ldots,r$.
If $\det\boldsymbol M\neq0$, we get that its rows $j$ have to be linearly independent, and, hence, the set $\{|\alpha_1\rangle,\ldots,|\alpha_r\rangle\}$ is linearly independent.
Let us additionally define the invertible diagonal matrices $\boldsymbol D_1={\rm diag}(\sqrt{0!},\ldots,\sqrt{(r-1)!})$ and $\boldsymbol D_2={\rm diag}(\exp[-|\alpha_1|^2/2],\ldots,\exp[-|\alpha_r|^2/2])$.
This yields the so-called Vandermonde matrix $\boldsymbol V=\boldsymbol D_2^{-1}\boldsymbol M\boldsymbol D_1$~\cite{LinAlg},
\begin{equation}
 V_{j,n}={\alpha_j}^n.
\end{equation}
Its determinant is given by
\begin{equation}
 \det \boldsymbol V=\prod_{1\leq i<j\leq r}(\alpha_i-\alpha_j)\neq 0,
 \label{eq:vandermonde}
\end{equation}
since $\alpha_i\neq\alpha_j$.
Due to the fact that $\boldsymbol D_1$ and $\boldsymbol D_2$ are invertible, $\det \boldsymbol V\neq 0$ is equivalent to $\det \boldsymbol M\neq 0$.
Hence, the coherent states $\{|\alpha_1\rangle,\ldots,|\alpha_r\rangle\}$ are linearly independent, and the output state in Eq.~\eqref{eq:BSexpansion} has the Schmidt rank of $r$.

We may further generalize the mapping in Eq.~\eqref{eq:BSexpansion} to the case of mixed states.
For this purpose we assume that the amplitudes of the input state obey the classical statistics $P_{\rm cl} (\{\alpha_i\},\{\kappa_i\})$, which is a joint probability for the coherent amplitudes $\alpha_i$ and arbitrary complex coefficients $\kappa_i$ with $i=1,\ldots,r$.
The corresponding mixed input state is transformed via
\begin{align}
 &\int {\rm d} P_{\rm cl} (\{\alpha_i\},\{\kappa_i\}) \sum_{i,j=1}^r \kappa_i \kappa_j^\ast |\alpha_i\rangle \langle \alpha_j| \otimes |0\rangle \langle 0|
 \nonumber\\ \,\stackrel{\rm BS}{\longmapsto}\,
 &\int {\rm d} P_{\rm cl} (\{\alpha_i\},\{\kappa_i\}) \sum_{i,j=1}^r \kappa_i \kappa_j^\ast \left|\frac{\alpha_i}{\sqrt 2},\frac{\alpha_i}{\sqrt 2}\right\rangle \left\langle\frac{\alpha_j}{\sqrt 2},\frac{\alpha_j}{\sqrt 2}\right |
\end{align}
into a mixed output state.
This is a classical mixture of pure states with a Schmidt rank less or equal to $r$, since, here, we allow some $\kappa_i$ to become zero in order to ensure the closure of the set of mixed states.

\subsection{Multipartite entanglement}
\begin{figure}[h]
	\includegraphics*[width=4.5cm]{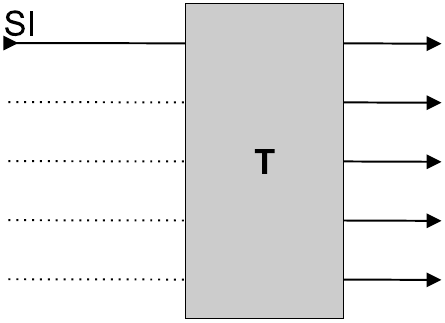}
	\caption{
	A nonclassical input state (SI) is combined with a classical $N-1$-mode vacuum state (dotted lines).
	In general, the output state shows multipartite entanglement.
	}\label{fig:Nsplitter}
\end{figure}

Let us further extend the approach to multipartite entanglement.
We may consider an $N$ splitter ($N$S) as given in Fig.~\ref{fig:Nsplitter}, being the generalization of a beam splitter to $N$ modes.
It acts on a multimode coherent state as
\begin{align}
	|\boldsymbol{\alpha}\rangle\stackrel{N\text{S}}{\longmapsto}|\boldsymbol T\boldsymbol\alpha\rangle,
\end{align}
for the unitary matrix $\boldsymbol T=(t_{j,j'})_{j,j'=1}^N$ and the coherent amplitudes $\boldsymbol\alpha\in\mathbb C^{N}$.
Let us note that this scheme is not restricted to an equal splitting of the input intensities into the output modes.
Especially in the bipartite case, $N=2$, this generalizes our previous results beyond the particular choice of a 50:50 beam splitter.

In this setup, the input state $|\Psi_{\rm Ncl}\rangle$ in Eq.~\eqref{eq:CS-dec} together with $N-1$ times vacuum transform as
\begin{align}
	|\Psi_{\rm Ncl},0,\ldots,0\rangle\stackrel{N\text{S}}{\longmapsto}\sum_{i=1}^{r} \kappa_i|t_{1,1}\alpha_i,\ldots,t_{N,1}\alpha_i\rangle.
\end{align}
For each output mode $j$ the states $\{|t_{j,1}\alpha_1\rangle,\ldots,|t_{j,1}\alpha_r\rangle\}$ are linearly independent for $t_{j,1}\neq0$; see Eq.~\eqref{eq:vandermonde}.
This follows from the fact that $\alpha_{i}\neq\alpha_{i'}$ is equivalent to $t_{j,1}\alpha_{i}\neq t_{j,1}\alpha_{i'}$.
Hence, the output state is a GHZ-type state; cf. Refs.~\cite{GHZ89,ZZH97,RLZL13}.
Such a state exhibits the same amount of genuine multipartite entanglement as the amount of nonclassicality of the input state, which follows from the generalized version of the Schmidt rank in Ref.~\cite{EB01}.

Our general finding shows that the resources, which arise from the amount of nonclassicality of a given input state, can be directly mapped onto the entanglement resource in the output ports of a beam splitter.
In this sense, the setup in Fig.~\ref{fig:setup} can be used in the following form.
A single-mode nonclassical state, having a nonclassicality quantified by $r$ superpositions of coherent states, can be prepared, e.g., by a nonlinear optical device.
A simple beam splitter maps the output ports to an entangled quantum resource with an amount of bipartite entanglement which is necessarily equal to $r$.
Replacing the beam splitter by an $N$ splitter, cf. Fig.~\ref{fig:Nsplitter}, one even gets genuine multipartite entanglement of strength $r$ from the single-mode nonclassical input state.
It is also worth mentioning that the replacement of the vacuum inputs by coherent states does not affect the amount of entanglement in the output states.

\subsection{Bipartite nonclassicality and entanglement correlations}
We started our consideration with the relation of single-mode nonclassicality to bipartite entanglement.
Now we generalize this approach to the notion to bipartite nonclassicality.
This property can be quantified by the number $R$ of superpositions of a bipartite coherent states,
\begin{align}
	|\Psi_{2-\rm Ncl}\rangle=\sum_{i=1}^{R} \kappa_i|\alpha_i,\beta_i\rangle,
	\label{eq:bi-ncl}
\end{align}
with $(\alpha_{i},\beta_{i})\neq (\alpha_{i'},\beta_{i'})$ for $i\neq i'$. At a sketchy look this property may appear to be closely related to entanglement; hence we need to consider it in more detail.

Let us consider the example of two copies of the state in Eq.~\eqref{eq:CS-dec}, which yields a bipartite product state $|\Psi_{\rm Ncl}\rangle^{\otimes 2}$ with a nonclassicality of $R=r^2$.
Hence the logarithm of $R$ is additive, $\log R=2\log r$, as it has been shown for the corresponding entanglement measure in Ref.~\cite{EB01}.
A 50:50 beam splitter transforms $|\alpha,\beta\rangle$ to $|(\alpha+\beta)/\sqrt 2,(\alpha-\beta)/\sqrt 2\rangle$.
These considerations can be used to get a brighter output state, as a first step towards so-called macroscopic entanglement~\cite{BMSSTG13,LGCPS13}.
Using the two copies of the single-mode nonclassical state as the two input modes of the beam splitter, the output state is
\begin{align}
	&(|\alpha\rangle-|-\alpha\rangle)^{\otimes 2}
	\\\nonumber\stackrel{\text{BS}}{\mapsto}\,&
	|\sqrt 2\alpha,0\rangle-|0,\sqrt 2\alpha\rangle-|0,-\sqrt 2\alpha\rangle+|-\sqrt 2\alpha,0\rangle
	\\\nonumber=&
	\left(|\sqrt 2\alpha\rangle+|-\sqrt 2\alpha\rangle\right)|0\rangle-|0\rangle\left(|\sqrt 2\alpha\rangle+|-\sqrt 2\alpha\rangle\right),
\end{align}
where the initial coherent amplitudes are scaled with $\sqrt 2>1$. The output state has the amount $R=4$ of bipartite nonclassicality.
The entanglement is, however, quantified by the Schmidt rank of $r=2$.

The state in Eq.~\eqref{eq:bi-ncl} can be entangled with a certain Schmidt rank or separable.
Moreover, a Schmidt number $R$ state needs an expansion of at least $R$ different coherent (product) states.
Hence, in this generalized scenario the bipartite nonclassicality is bounded from below by the Schmidt number.
Similarly, this bounding property is valid for multimode nonclassicality and entanglement.

We finally note that nonclassicality has been also considered in the context of general time-dependent quantum correlations of light~\cite{V08}.
Using the fact that the amount of single-mode input nonclassicality is perfectly mapped to the amount of output entanglement, it is possible to define time-dependent entanglement correlations.
For example, the scenario in Eq.~\eqref{eq:ent-photon} together with a time delay in one output mode could be interpreted as quantum entanglement between different times.
The most elementary scenario of this type is the well-known photon antibunching effect~\cite{KDM77}.
In this case the single-mode input state is replaced by a field showing nonclassical time-delayed intensity correlations, which violate a Schwarz inequality.
In the output ports of the beam splitter this yields a kind of time-delayed entanglement.

\section{Summary and conclusions}\label{sec:sc}
We have considered the relation of nonclassicality of a single-mode radiation field and the entanglement, which is obtained by splitting the nonclassical state by a beam splitter.
The quantification of both properties, nonclassicality and entanglement, is strictly based on the number of quantum superpositions of classical states with respect to the corresponding quantum property.
We have shown that the amount of single-mode nonclassicality of the input state directly maps to the same amount of entanglement available in the two output ports of the beam splitter.
Using a multiport splitter, one can even map the amount of single-mode nonclassicality onto the same amount of multipartite entanglement.
We additionally showed that the amount of multipartite nonclassicality is an upper bound for the entanglement.
We also outlined the possibility to relate time-dependent quantum correlations to entanglement between different times.

Altogether, this opens a variety of possibilities to properly prepare the nonclassical single-mode state, in order to obtain a desired class of entangled quantum states.
Often it is much simpler to prepare a single-mode state than an entangled one.
On this basis it is obvious that the nonclassicality of the single mode under control may be considered as the resource for such applications in quantum technology, which are usually based on the property entanglement.
It is worth noting that our approach not only applies to radiation fields but also to other harmonic degrees of freedom, such as the quantized motion of trapped atoms and others.

\section*{Acknowledgments}
This work was supported by the Deutsche Forschungsgemeinschaft through SFB 652.


\begin{thebibliography}{99}
\bibitem{glauber1} R. J. Glauber, Phys. Rev. {\bf 130}, 2529 (1963).
\bibitem{glauber2} R. J. Glauber, Phys. Rev. {\bf 131}, 2766 (1963).
\bibitem{glauber3} R. J. Glauber, Phys. Rev. Lett. {\bf 10}, 84 (1963).
\bibitem{klaud-sud} J. R. Klauder and E. C. G. Sudarshan, {\it Fundamentals of Quantum Optics} (W. A. Benjamin, New York, 1968).
\bibitem{tit-glau} U. M. Titulaer and R. J. Glauber, Phys. Rev. {\bf 140}, B676 (1965).
\bibitem{mandel} L. Mandel, Phys. Scr. {\bf T12}, 34 (1986).
\bibitem{ma-wo} L. Mandel and E. Wolf, {\it Optical Coherence and Quantum Optics} (Cambridge University Press, Cambridge, 1995).
\bibitem{vo-we} W. Vogel and D.-G. Welsch, {\it Quantum Optics}, 3rd ed. (Wiley-VCH, Weinheim, 2006).
\bibitem{agar} G. S. Agarwal, {\it Quantum Optics} (Cambridge University Press, Cambridge, 2013).
\bibitem{sud} E.C.G. Sudarshan, Phys. Rev. Lett. {\bf 10}, 277 (1963).
\bibitem{epr} A. Einstein, N. Rosen, and B. Podolsky, Phys. Rev. \textbf{47}, 777 (1935).
\bibitem{schroed} E. Schr\"odinger, Naturwissenschaften \textbf{23}, 807 (1935).
\bibitem{nielsenchuang} M. A. Nielsen and I. L. Chuang, {\it Quantum Computation and Quantum Information} (Cambridge University Press, Cambridge, 2000).
\bibitem{horodecki} R. Horodecki, P. Horodecki, M. Horodecki, and K. Horodecki, Rev. Mod. Phys. {\bf 81}, 865 (2009).
\bibitem{guehne} O. G\"uhne and G. T\'oth, Phys. Rep. {\bf 474}, 1 (2009).
\bibitem{werner} R. F. Werner, Phys. Rev. A {\bf 40}, 4277 (1989).
\bibitem{sanpera} A. Sanpera, R. Tarrach, and G. Vidal, Phys. Rev. A {\bf 58}, 826 (1998).
\bibitem{spe-vo2} J. Sperling and W. Vogel, Phys. Rev. A {\bf 79}, 042337 (2009).
\bibitem{Raymond} J. M. Raimond, M. Brune, and S. Haroche, Rev. Mod. Phys. {\bf 73}, 565 (2001).
\bibitem{ahar} Y. Aharonov, D. Falkoff, E. Lerner, and H. Pendleton, Ann. Phys. (N.Y.) {\bf 39}, 498 (1966).
\bibitem{kim} M. S. Kim, W. Son, V. Buz\v{e}k, and P. L. Knight, Phys. Rev. A {\bf 65}, 032323 (2002).
\bibitem{xiangbin} Wang Xiang-bin, Phys. Rev. A {\bf 66}, 024303 (2002).
\bibitem{WEP03} M. M. Wolf, J. Eisert, and M. B. Plenio, Phys. Rev. Lett. {\bf 90}, 047904 (2003).
\bibitem{JLC13} Z. Jiang, M. D. Lang, and C. M. Caves, Phys. Rev. A {\bf 88}, 044301 (2013).
\bibitem{asboth} J. K. Asb\'oth, J. Calsamiglia, and H. Ritsch, Phys. Rev. Lett. {\bf 94}, 173602 (2005).
\bibitem{JJS09} D. Jian, T. Jian-Fu, and W. Shu-Yan, Commun. Theor. Phys. {\bf 51}, 823 (2009).
\bibitem{Z11} P. Zawadzki, Quantum Inf. Process. {\bf 11}, 1419 (2012).
\bibitem{gehrke} C. Gehrke, J. Sperling, and W.Vogel, Phys. Rev. A {\bf 86}, 052118 (2012).
\bibitem{uhlmann} A. Uhlmann, Open Syst. Inf. Dyn. {\bf 5}, 209 (1998).
\bibitem{spe-vo1} J. Sperling and W. Vogel, Phys. Scr. {\bf 83}, 045002 (2011).
\bibitem{dodonov}  V. V. Dodonov, I. A. Malkin, and V. I. Manko, Physica (Amsterdam) {\bf 72}, 597 (1974).
\bibitem{HBBuch} K. Yosida, {\it Functional Analysis} (Springer, Berlin, 2008)
\bibitem{LinAlg} R. A. Horn and C. R. Johnson, {\it Topics in Matrix Analysis} (Cambridge University Press, Cambridge, 1991).
\bibitem{EB01} J. Eisert and H. J. Briegel, Phys. Rev. A {\bf 64}, 022306 (2001).
\bibitem{GHZ89} D. M. Greenberger, M. A. Horne, and A. Zeilinger, {\it Bell’s Theorem, Quantum Theory, and Conceptions of the Universe} (Kluwer, Dordrecht, 1989).
\bibitem{ZZH97} M. \.{Z}ukowski, A. Zeilinger, M. A. Horne, Phys. Rev. A {\bf 55}, 2564 (1997).
\bibitem{RLZL13} J. Ryu, C. Lee, M. \.{Z}ukowski, and J. Lee, Phys. Rev. A {\bf 88}, 042101 (2013).
\bibitem{BMSSTG13} N. Bruno, A. Martin, P. Sekatski, N. Sangouard, R. T. Thew, and N. Gisin, Nat. Phys. {\bf 9}, 545 (2013).
\bibitem{LGCPS13} A. I. Lvovsky, R. Ghobadi, A. Chandra, A. S. Prasad, and C. Simon, Nat. Phys. {\bf 9}, 541 (2013).
\bibitem{V08} W. Vogel, Phys. Rev. Lett. {\bf 100}, 013605 (2008).
\bibitem{KDM77} H. J. Kimble, M. Dagenais, and L. Mandel, Phys. Rev. Lett. {\bf 39}, 691 (1977).
\end{thebibliography}
\end{document}